\documentclass[manuscript]{aastex}

\def\gsim{\mathrel{\raise.5ex\hbox{$>$}\mkern-14mu
             \lower0.6ex\hbox{$\sim$}}}
\def\lsim{\mathrel{\raise.3ex\hbox{$<$}\mkern-14mu
             \lower0.6ex\hbox{$\sim$}}}
\def\simless{\mathbin{\lower 3pt\hbox
     {$\rlap{\raise 5pt\hbox{$\char'074$}}\mathchar"7218$}}}   
\def\simmore{\mathbin{\lower 3pt\hbox
     {$\rlap{\raise 5pt\hbox{$\char'076$}}\mathchar"7218$}}}   

\shorttitle{Luminosity Correlations for Gamma-Ray Bursts}
\shortauthors{Sultana et al.}

\begin{document}

\title{The supercritical pile gamma-ray burst model:
The GRB afterglow steep-decline-and-plateau phase}

\author{J. Sultana\altaffilmark{1}, D. Kazanas\altaffilmark{2} and A. Mastichiadis\altaffilmark{3}}

\email{joseph.sultana@um.edu.mt}

\baselineskip= 14pt

\altaffiltext{1}{Mathematics Department, Faculty of Science,
University of Malta, Msida MSD2080 Malta.}
\altaffiltext{2}{Astrophysics Science Division, NASA/Goddard Space
Flight Center, Greenbelt, MD 20771 USA.}
\altaffiltext{3}{Department of Physics, University of Athens, Panepistimiopolis, GR 15783, Zografos, Greece}

\begin{abstract}

We present a process that accounts for the steep-decline-and-plateau
phase of the Swift-XRT light curves, vexing features of GRB
phenomenology. This process is an integral part of the
``supercritical pile" GRB model, proposed a few years ago to provide
an account for the conversion of the GRB kinetic energy into
radiation with a spectral peak at $E_{\rm pk} \sim m_ec^2$. We
compute the evolution of the relativistic blast wave (RBW) Lorentz
factor $\Gamma$ to show that the radiation--reaction force due to
the GRB emission can produce an abrupt, small ($\sim 25\%$) decrease
in $\Gamma$ at a radius which is smaller (depending on conditions)
than the deceleration radius $R_D$. Because of this reduction, the
kinematic criticality criterion of the ``supercritical pile" is no
longer fulfilled. Transfer of the proton energy into electrons
ceases, and the GRB enters abruptly the afterglow phase at a
luminosity smaller by $\sim m_p/m_e$ than that of the prompt
emission. If the radius at which this slow-down occurs is
significantly smaller than $R_D$, the RBW internal energy continues
to drive the RBW expansion at a constant (new) $\Gamma$, and its
X-ray luminosity remains constant until $R_D$ is reached, at which
point it resumes its more conventional decay, thereby completing the
``unexpected" XRT light curve phase. If this transition occurs at $R
\simeq R_D$, the steep decline is followed by a flux decrease
instead of a ``plateau", consistent with the conventional afterglow
declines. Besides providing an account of these peculiarities, the
model suggests that the afterglow phase may in fact begin before the
RBW reaches $R \simeq R_D$, thus introducing novel insights into the
GRB phenomenology.

\end{abstract}

\keywords{cosmological parameters --- Gamma-ray burst: general}

\section{Introduction}

Gamma-ray Bursts (GRBs) are extremely bright explosions at
cosmological  distances \citep{cos97,par97}, with isotropic
luminosities occasionally exceeding $\sim 10^{54}$ erg/sec. Their
durations are in the range $\sim 0.1 - 1000$ sec, and their
luminosity peaks at an energy close to the electron rest mass
energy, $E_{\mathrm{pk}} \sim 1$ MeV  {(however, the accumulation of
observational data has shown that this characteristic energy
exhibits a wider distribution, ranging from as low as a few keV
\citep{campana06} to as high as $15$ MeV \citep{axelsson12}, in
correlation with either the isotropic energy released in the burst,
$E_{\rm iso}$, or its peak isotropic luminosity $L_{\rm p, iso}$.})

They are believed to originate in the collapse of stellar cores
(long GRBs) or the mergers of neutron stars (short GRBs), processes
which result in jet-like relativistic outflows of Lorentz factors
$\Gamma \lsim 300$ (but on occasion exceeding values 500 -1000
\citep{abdo09a, abdo09b, ackermann10,has12}). It is generally
considered that the kinetic energy of these outflows is converted
efficiently into radiation in collisions of shells of matter ejected
at different times by the GRB ``central engine"; such collisions are
thought necessary in order to produce the observed rapid GRB
variability \citep[see however][]{NK09} and spectra with the
characteristic GRB signature, i.e. maximum luminosity at energies
$E_{\rm pk} \simeq m_{e} c^2$ (for reviews see \citet{pir04} for
bursts prior to the launch of {\em Swift} and \citet{Z07} for bursts
post the {\em Swift} launch). Following this most luminous, prompt,
$\gamma-$ray emission phase, GRBs shift into their afterglow phase.
In this phase their luminosity is substantially lower, and their
peak emission shifts into the X-ray band. The longer duration of
this phase ($\Delta t \sim 10^5$ s) allows their more precise
localization and optical detection, which can then provide their
redshift.

According to prevailing theory \citep{pir04,Z07} GRB emission is due
to synchrotron radiation by electrons accelerated, in the prompt
phase, in the shocks of the colliding shells, while in the afterglow
in the forward shock of the expanding RBW. As the RBW expands it
sweeps more matter and after it has swept-up an amount $M \simeq
E/c^2\Gamma^2$ ($E$ is its total injected energy, $\Gamma$ its
asymptotic Lorentz factor) at a distance $R_D$, its Lorentz factor
decreases and it is thought to enter the afterglow phase, as
surmised by the declining X-ray and optical fluxes. Under these
assumptions one can calculate the expected X-ray flux decrease with
time, which turns out to be a power law, $F_X \propto t^{-\alpha}$,
with $\alpha \simeq 1$ in spherical \citep{sar98} and $\alpha \simeq
2$ in jet-like \citep{sar99} flows. Indeed the early, sparsely
sampled, pre-{\em Swift} light curves appeared consistent with such
a behavior. However, their more densely sampled X-ray light curves
with the XRT aboard {\em Swift} uncovered significant deviations
from this behavior. So, following the prompt $Swift-$BAT
$\gamma-$ray emission, typical XRT afterglows
\citep{tagliaferri05,nousek,evans09} comprise a segment of much
steeper flux decline ($\propto t^{-3}\ \mbox{to}\ t^{-6}$), followed
by either a less steep power law \citep{liang07}, or a
$10^2 - 10^5$ sec period of nearly constant flux (a ``plateau"),
followed finally at $t = T_{\rm brk}$ by the more conventional
power-law decline $\simeq t^{-1}$. In addition, {\em Swift}
follow-ups discovered \citep{burrows05} also occasional
flares on top of these light curves, as late as $\sim 10^5$ sec
since the BAT trigger. These unexpected details in the GRB afterglow
light curves were added to the other already open problems related
to the GRB prompt emission, namely the nature of their ``inner
engine'', the non-dissipative transport of their energy to the
emission region to distances $R\sim 10^{16}-10^{17} \mathrm{cm}$,
its efficient dissipation there and the physics behind the
distribution of the GRB peak energy $E_{\mathrm{pk}} \sim 1
\mathrm{MeV}$ \citep{mal95,pre00}.

{ These prompt GRB emission issues are usually settled  by {\em
fiat} in most literature, e.g. by assuming that a large fraction
($\sim 50\%$) of the proton post-shock energy is converted into
electrons with a minimum Lorentz factor $\gamma_{min}$, chosen so
such that the burst luminosity maximum would appear at $E_{\rm pk}$.
However, this exhausts essentially most of the models' freedom,
setting the afterglow evolution on the path described in
\citet{sar98}, thus making an account of the observed afterglow
light curves shape all the more pressing. There have been a number
of attempts to account for at least some of these features. Thus,
\citet{kaz07} suggested that if the post-shock proton distribution
function comprises, in addition to a relativistic Maxwellian of $T
\sim \Gamma m_pc^2$, also a power law that extends to energies
$\gamma m_p \gg \Gamma m_p$,
one could account for the steep decline followed by the more
conventional $\propto t^{-1}$ power law decline light curves of GRB
afterglows. A similar proposal was put forward by \citet{GS09} who
employed the realistic electron distributions produced in PIC
simulations, which do have a form similar to that conjectured in
\citet{kaz07}. More recently, \citet{PMP11} provided models with the
desired general afterglow shape by adjusting the maximum electron
distribution Lorentz factor $\gamma_{max}$ in such a way that the
steep decline represents synchrotron emission by electrons near
$\gamma_{max}$ (which are cooling fast), while attributing the
constant X-ray flux component to inverse Compton emission by the
(much slower varying) lower energy section of the electron
distribution. {Another commonly accepted interpretation of the steep
decay phase in the GRB early afterglow light curves is due to the
high-latitude emission, i.e. the curvature effect \citep[][and
references therein]{zhang06} that follows the prompt emission phase.
The goal of of the present note is to indicate that the XRT
afterglow light curves can be accounted for by incorporating the
supercritical pile model (hereafter SPM) and its feedback on the
dynamics of the GRB relativistic blast wave (RBW) that gives rise to
the GRB. } }

The SPM \citep{kaz02,mas06,mas09} has been introduced to address the
issue of the GRB dissipation and the apparently efficient conversion
of the RBW kinetic energy into radiation with the observed spectral
characteristics. The fundamental process of this model is a
radiative instability which can convert the internal energy of the
RBW relativistic protons into relativistic $e^{+}\,e^{-}$ pairs. The
conversion takes place on timescales $\Delta R/c$~ ($\Delta R$ is
the typical width of the RBW), via the $p\,\gamma\rightarrow
p\,e^{+}\,e^{-}$ reaction, provided that certain kinematic and
dynamic threshold (criticality) conditions are fulfilled, which are
discussed in the next section.
%
%
Unlike the more conventional GRB models in the literature, the SPM
does not require (but does not forbid) accelerated particle
populations besides those produced by the isotropization of the RBW
kinetic energy behind the shock. Most importantly, it does
not invoke an {\em ad hoc} equipartition between the proton and
electron energy densities behind the shock, as it includes
the dynamics that convert the proton energy into $e^+e^-$. Finally,
a consequence of the kinematic threshold condition of the
$p\,\gamma\rightarrow p\,e^{+}\,e^{-}$ reaction, is the natural
emergence (after all relevant Lorentz transformations) of a
characteristic photon peak energy, $E_{\rm pk} \simeq 1~\mathrm{MeV}$ (in the
observer frame, assuming that the process operates close to
its kinematic threshold), largely independent of the RBW Lorentz
factor and in agreement with observations \citep[e.g][]{mal95,gol12}. In
more conventional models, such a characteristic photon energy occurs
only at the expense of assuming the presence of a low energy cut-off
in the electron distribution function.


More recently \citet{mas09} explored numerically the SPM  from the
prompt into the afterglow stage by computing the evolution and
spectrum of a RBW of Lorentz factor $\Gamma_0 = 100$ propagating
through a medium of density $n(r) \propto r^{-2}$, representing the
wind of a Wolf-Rayet star. In this treatment they incorporated in
the RBW evolution equations the effects of the radiative drag
introduced in the production of the GRB spectra by the bulk
Comptonization of upstream scattered photons, a crucial element of
this model. They showed that when both threshold conditions of the
model were fulfilled and the energy stored in protons is converted
into radiation, the resulting radiation reaction force reduced the
RBW Lorentz factor $\Gamma$ and also the GRB flux. { In the
particular case they examined, the drop in $\Gamma$ was sufficiently
large to render the RBW subcritical. The GRB thus entered the
afterglow phase (with its luminosity coming only from the electrons
being swept-up by the RBW) after only a couple of seconds (in the
observer frame) thus producing a short GRB, even though the RBW was
assumed to propagate through a presupernova stellar wind medium. }

In this work we employ a simplified version of the RBW evolution
discussed in \citet{mas09} to concentrate our attention to the
entire evolution of  $\Gamma$ from its acceleration phase into its
late time decline. Using this simplified version we provide an
account of the vexing steep-decline-and-plateau phase in the early
GRB afterglow light curves observed in most long GRBs. The
simplified version of this approach allows us to compute the
evolution of both conical and parabolic GRB jet configurations in a
medium of constant density $\rho$ from their accelerating phase
(i.e. $\Gamma \simeq 1$) to their adiabatic decay past the
deceleration radius. In Section 2 we introduce the general framework
of the supercritical pile model and criticality conditions. In
Section 3 we present our results and then in Section 4 we summarize
our findings and present our conclusions.

\section{The ``Supercritical Pile'' Model in Brief}

The process described in this section was first used in the context
of active galactic nuclei by \citet{kaz99} who employed it to argue
for the possibility of a hadronic origin of the relativistic
electrons in blazars. It involves the combination of: 1. The
relativistic proton plasma radiative instability of \citet{kir92} as
applied to a RBW.  2. The increase in the energy of synchrotron
photons produced in this plasma, which, upon their scattering in
upstream located matter (referred to as the ``mirror") are then
re-intercepted by the RBW, as discussed in \citep{ghi96}. The
instability of \citet{kir92} is basically that of a nuclear pile:
Synchrotron photons produced by $e^+e^-$ pairs, interact with the
relativistic protons of the plasma to produce more $e^+e^-$ pairs;
the set up is radiatively unstable if the column of the plasma is
sufficiently large that at least one of the $N$ photons ($N \simeq
\gamma/b \gamma^2 = 1/b \gamma$; $\gamma$ is the electron Lorentz
factor and $b$ the magnetic field normalized to the critical one
$B_c = m_e^2 c^3/e \hbar \simeq 4.4 \times 10^{13}$ G) produced by a
member of an $e^+e^-$ pair of energy $E_e \simeq \gamma m_ec^2$,
produces another pair in reaction with a relativistic proton before
escaping the system by the process $p\gamma\rightarrow
p\,e^{-}\,e^{+}$.

If $R$ is the size of the plasma and $n_0$ the proton density, this
last constraint reads $\sigma_{ p \gamma} Rn_{0}  \gsim 1/N \simeq
b\gamma$. However, the $p \gamma$ process requires that the energy
of the synchrotron photon $E_s$ be sufficiently high to produce a
pair on the proton rest frame, i.e. $\gamma \, E_s \gsim 2m_ec^2$.
Considering that $E_s \simeq b \gamma^2 m_ec^2$ the kinematic
threshold reads $\gamma^3 \gsim 2/b$. Incorporating this in the
column density constraint one gets $\sigma Rn_{0} \gamma^2 \gsim 2$.
Applying these considerations to the particles in the postshock
region of a RBW one can set their $\gamma$ equal to the RBW Lorentz
factor $\Gamma$, so that the criticality conditions are expressed in
terms of the bulk RBW kinematic properties, i.e. $\Gamma^3 \gsim
2/b$ and $\sigma_{p \gamma} Rn_{0} \Gamma^2 \gsim 2$.


If the synchrotron photons scatter upstream of the RBW in a
``mirror" (in AGN this mirror are the BLR clouds) of scattering
depth $\tau_{\rm mirr}$, upon their re-interception by the RBW they
have energies larger by $4 \Gamma^2$. This modifies the kinematics
and also the column density conditions to
\begin{equation}
\Gamma^5 \gsim (1/2b)  ~~~~~ {\rm and } ~~~~~ 2\tau_{\mathrm{mirr}}n_{0}\sigma R\Gamma^{4} \geq 1 \label{kc}
\end{equation}
With the above setting for the conversion of proton energy into
electrons and photons, it was shown in \citet{kaz02}, that the
threshold of the $p\gamma\rightarrow p\,e^{-}\,e^{+}$ reaction,
translates on the observer's frame to an energy $b \Gamma^5$, which
by the first of the relations above implies a peak emission energy
at roughly the electron rest mass, in agreement with observations.

The observed peak energy occurs at this value only if the process
operates close to the kinematic threshold at all time. The fact that
$b\Gamma^5 \sim 1/2$ does not suffice to produce a burst, because,
while $b\Gamma^5$ may be well above the threshold, say $10$ MeV or
higher, rapid proton energy release requires also that the column of
swept up protons be sufficiently high such that the dynamic
threshold in (\ref{kc}) be also satisfied. Apparently, the larger
$b\Gamma^5$ is, the faster the accumulated energy will be released
once supercritical. Therefore the model does not exclude the
higher values of $E_{pk}$ observed recently
\citep{axelsson12,Guiriec13}. More importantly, the bright bursts
analyzed in these references indicated a correlation between the
value of $E_{pk}$ and the burst luminosity ($L \propto
E_{pk}^{\alpha}, ~\alpha \simeq 1.3$) during the evolution of
the same burst (not addressed to the best of our knowledge by to
date models), suggesting additional nuances, on which it would be
unwise to speculate at this point. 

On the other hand (averaged over the burst duration) values of
$E_{pk}$ smaller than $1$ MeV \citep{campana06} may be due to larger
viewing angles of the GRB jet $\theta > 1/\Gamma$  as discussed in
\citet{ioka01}. Also, $E_{pk} < 1$ can be obtained for $\theta <
1/\Gamma$, if the shock produces in addition to protons with energy
$\sim\Gamma m_p c^2$, also a power law tail that extends to energies
$\gamma \gg \gamma_1
> \Gamma$ such that $b \gamma_1^5 \gsim 1/2$, while $b\Gamma^5 <
1/2$,  as discussed in \citet{kaz07}. Independently of the
specifics discussed above, the important point to bear in mind is
the emergence of a characteristic energy in agreement with
observations, after all Lorentz transformations have taken place, as
a result of the physics of the dissipation process.


The evolution of $\Gamma$ of a RBW is given by the coupled mass and
energy-momentum conservation laws \citep[see e.g.][]{chi99}. In case
that the RBW plows through a radiation field, one must also include
the effects of radiation reaction of the RBW as it plows through the
radiation that has scattered upstream of the shock
\citep{mas08,boe09}; these are given below
%
%
%
\begin{equation}
\frac{dM}{dR} = 4\pi R^2 \Gamma \rho - \frac{\dot{E}}{c^3\Gamma}, \label{evo1}
\end{equation}
and
\begin{equation}
\frac{d\Gamma}{dR} = \Gamma\frac{AR_{0}/R^2}{1 + AR_{0}/R} - \frac{4\pi R^2\rho\Gamma^2}{M} -
\frac{F_{\mathrm{rad}}}{{Mc^2}}, \label{evo2}
\end{equation}
where $R(t)$ is the radius measured from the center of the original
explosion, $A = E_{0}/M_{0}c^2$ with $E_{0},\  M_{0}$ representing
the initial total energy and rest mass respectively of the flow at
$R = R_{0}$, corresponding to the radius of the GRB progenitor. Here
$\dot{E}$ represents the radiation emission rate as measured in the
comoving frame and $F_{\mathrm{rad}}$ is the radiation reaction
force exerted on the RBW by the radiation field exterior to the
flow; this is given by
\begin{equation}
F_{\rm rad} = \frac{64\pi}{9c}\tau_b
n_{e}\sigma_{T}R\Gamma^4\dot{E},
\end{equation}
where $\tau_{b}$ is the RBW Thomson depth, $\sigma_{T}$ is the
Thomson cross-section and $n_{e}$ is the CSM electron density
assumed to be the same as the proton density $n_0$ used in Eq.
(\ref{kc}).

\citet{mas09} applied these equations to compute both the evolution
of  the Lorentz factor $\Gamma$ and the emitted radiation for a RBW
propagating through the wind of a Wolf-Rayet star, i.e. a medium
with density profile $n(R) \propto R^{-2}$, that presumably being
the progenitor of a supernova that gave rise to the RBW.  The values
of  $\dot{E}$ and $F_{\rm rad}$ were computed by implementing the
numerical code, originally described in \citet{mas06}, to solve the
equations
\begin{equation}
\frac{\partial n_{i}}{\partial t} + L_{i} + Q_{i} = 0, \label{ke}
\end{equation}
where the functions $n_{i}$ represent the differential number
densities of protons, electrons and photons with the index $i$
taking any of the subscripts $p$, $e$ or $\gamma$, while $L_{i}$
denotes the losses and escape, and $Q_{i}$ denotes the injection and
source terms in the system. 

The detailed calculations of \citet{mas09}, using parameters $R_{0} =
10^{14}\,{\rm cm}$, $n_{0} = 8\times10^{8}\,{\rm cm}^{-3}$,
$\Gamma_{0} = 100$, $B_{0} = 4.4\times104\,\mathrm{G}$, and $E_{0} =
10^{54}\,{\rm erg}$ different from those used herein -- except for
$E_0$ -- but appropriate for the setting considered, confirmed the
qualitative estimates concerning the positions and effects of the
problem thresholds. In addition, they showed the radiation reaction
effects to be significant, having an immediate effect on $\Gamma$,
which slowed down over a length scale short compared to $R_0$, to a
value lower than that required by the kinematic threshold of the
problem. This resulted in the precipitous decline of the GRB
luminosity as the only available energy to be radiated from that
point on was that of the swept-up electrons. At the same time they
computed the resulting spectrum, found in agreement with the basic
premises of the SPM (i.e. exhibiting a peak at $E_{\rm pk} \sim
m_ec^2$), with the spectrum softening significantly with the
decrease in $\Gamma$ effected by the radiation reaction process,
with the GRB thus entering the afterglow stage.


\section{Results}

In the present paper we study within the SPM a simplified
version of the evolution of the RBW Lorentz factor $\Gamma$ in a
medium of constant number density $n_{0}$; to this end, we begin our
computations at the radius of the GRB progenitor, $R_{0}$, where we
set $\Gamma_{0} = 1$, with approximate estimates for the resulting
luminosity, a fact that allows a much broader search in parameter
space and exploration of the evolution over longer time scales. The
evolution of $\Gamma$ is followed from its initial
accelerating phase, to its saturation (constant $\Gamma$) and
slow-down stages, attributing each to the prompt or afterglow stage
depending on whether the criticality conditions are fulfilled.

To reduce the number of free parameters we assume that the magnetic
field is in equipartition with the post-shock pressure so that
$B\simeq 0.4(n_{0}/1{\rm cm}^{-3})^{1/2}\Gamma$ Gauss. Then the
long-term evolution of the Lorentz factor depends on the free
parameters $E_{0}$, $R_{0}$ and $n_{0}$ which determine the radius
where the kinematic and dynamic conditions are satisfied so that the
RBW becomes supercritical. It could happen that for certain
parameter combinations the threshold conditions are satisfied at
more than one radius, in which case the released energy
should be proportional to the time between such bursts, as
found by \citet{ram01}. We examine both conic and parabolic
configurations of the GRB jet with parameters $R_{0} = 10^{11}\,{\rm
cm}$, $n_{0} = 100\,{\rm cm}^{-3}$, $M_{0}c^2 = 5\times
10^{51}\,{\rm erg}$, and total isotropic energy $E_{0} =
10^{54}\,{\rm erg}$.

\subsection{Conic Outflows}

\begin{figure}
\includegraphics[scale=1.5]{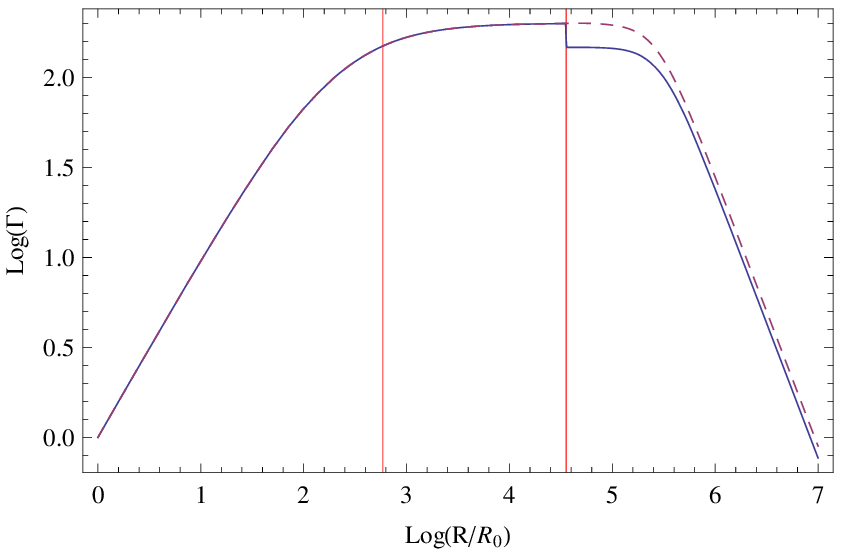}
\caption{The Lorentz factor for a RBW propagating in a constant
density environment for a conic GRB outflow. The dashed curve
represents the evolution with radius in the adiabatic case without
radiative drag. The red vertical lines represent the region where
the kinematic condition is satisfied, and the green line gives the
radius at which the dynamic condition starts to be satisfied. The
GRB parameters are given in the text. \label{fig1}}
\end{figure}

The evolution of $\Gamma$ is given by the solution of the coupled
equations (\ref{evo1}) and (\ref{evo2}). To simplify our treatment,
instead of using the numerical code employed in \citet{mas06} and
\citet{mas09} to calculate the radiation emission rate $\dot{E}$, we
use the fact that once the criticality conditions are satisfied
almost all the energy in the swept-up protons is immediately
radiated away, so that
\begin{equation}
\dot{E} = \dot{E}_{\rm inj} = 4\pi R^2\rho(\Gamma^2 - \Gamma)c^3,
\end{equation}
where $\dot{E}_{\rm inj}$ is the proton energy injection rate
\citep{bla76}. Letting $x = R/R_{0}$ the evolution equations in
(\ref{evo1}) and (\ref{evo2}) become
\begin{equation}
\frac{dM}{dx} = 4\pi R_{0}^3 x^2 \Gamma\rho - 4\pi
x^2R_{0}^3\rho(\Gamma - 1), \label{evoc1}
\end{equation}
and
\begin{equation}
\frac{d\Gamma}{dx} = \frac{\Gamma A}{x(x + A)} - \frac{4\pi x^2
R_{0}^3 \rho \Gamma^2}{M} -
\frac{256\pi^2}{9M}\tau_{b}n_{e}\sigma_{T}x^3
R_{0}^4\rho\Gamma^4(\Gamma^2 - \Gamma). \label{evoc2}
\end{equation}
As long as the threshold conditions in (\ref{kc}) are not satisfied
then the evolution of the RBW is described approximately by
(\ref{evoc1}) and (\ref{evoc2}) without the $\dot{E}$ and $F_{\rm
rad}$ terms on the RHS. This is the standard non-radiative case
mentioned earlier in which the Lorentz factor reaches its asymptotic
value $\Gamma \simeq A = E_{0}/M_{0}c^2 = 200$ and then proceeds
with the conventional decline of afterglow theory,  shown by the
dashed curve in Figure \ref{fig1}. For the chosen values of the GRB
parameters, the kinematic condition is satisfied at
$\log(R/R_{0})\simeq2.77$ (represented by the first vertical red
line in Figure \ref{fig1}) before the Lorentz factor reaches even
its asymptotic value. Once enough matter is piled up such that the
column of accumulated hot protons exceeds its critical value, the
RBW becomes supercritical at $\log(R/R_{0})\simeq4.55$ (represented
by the green vertical line in Figure \ref{fig1});  the proton
accumulated energy is released on the shock light crossing time
scale and results in a sudden drop in $\Gamma$ of the RBW due to the
radiative drag. The decrease in $\Gamma$ reduces the value of
$b\Gamma^5$ below its threshold value at $\log(R/R_{0})\simeq4.56$
(represented by the second red line in the figure) and arrests the
conversion of proton energy into radiation. The luminosity drops
precipitously, by roughly a factor $m_p/m_e$ ($m_p, m_e$ are the
proton and electron masses respectively) as the emitted radiation
now comes only from the cooling of the electrons swept by the RBW,
and the GRB enters the afterglow stage (one should note here that,
in distinction to most models, the SPM provides a natural, physical
grounds separation of  GRBs in prompt and afterglow stages).

However, despite the decrease in luminosity that ensues the
reduction in $\Gamma$ due to the effects of radiation reaction, the
rest mass accumulated to this point maybe too small, for the given
RBW internal energy $E_0$, than necessary to produce a decrease in
$\Gamma$ in the manner expected generally for distances $R> R_D$.
Therefore, the RBW evolution has no choice but continue at a
constant (or even increasing) $\Gamma$, even though, according to
the premises of the SPM, the GRB has entered the afterglow stage
(the kinematic threshold condition is not satisfied, the $p \gamma
\rightarrow p \, e^+ e^-$ reaction does not take place and hence
$E_{\rm pk} \ll m_e c^2$). The value of $\Gamma$ will remain
constant to a distance equal to the deceleration radius $R_D$, as
shown in Figure \ref{fig1}. We contend that this stage is
responsible for the ``plateau" observed in the XRT light curves.

In Figures 3a, 3b we plot the evolution of two RBWs with the same
initial conditions as those of Figures 1 and 2, except for the value
of $M_0 c^2$, which now determines the asymptotic value of
$\Gamma_{\infty}$, in the absence of radiation reaction. The value
of $M_0 c^2$ for these features is set to $M_0 c^2 = 3.5 \times
10^{51}$ erg and $M_0 c^2 = 2 \times 10^{51}$ erg for Figures 3a, 3b
respectively, implying asymptotic $\Gamma$ values of $\simeq 286
~{\rm and} ~ 500$. For these conditions, the radiation reaction
feedback is much larger, with the sharp transition taking place at a
distance increasingly closer to $R_D$ with increasing value of
$\Gamma_{\infty}$. This then implies a concomitant decrease in the
length of the constant $\Gamma$ section in the afterglow stage, the
latter effectively disappearing for the largest value of
$\Gamma_{\infty}$. This is of interest because of the correlation
between the ``plateau" luminosity and the time $T_{\rm bk}$ of its
break to the more conventional afterglow decrease with time
\citep{dai08,dai10}. It is worth noting that the evolution of
$\Gamma$ of Fig. 3b would likely correspond to one of the typical
shapes of the afterglow curves indicated in \citet{WillOBr}, i.e.
that of a steep decline, followed by a conventional time decrease of
the XRT flux.


\subsection{Parabolic Outflows}

\begin{figure}
\includegraphics[scale=1.5]{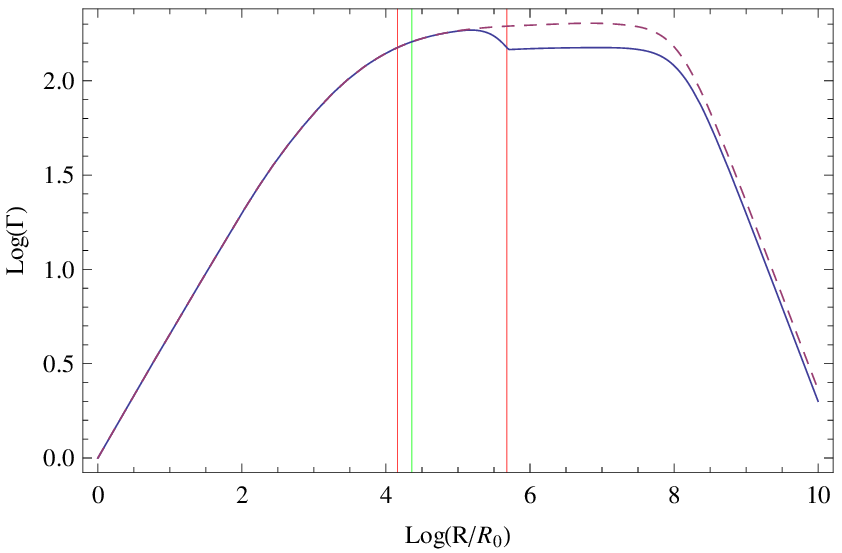}
\caption{The Lorentz factor for a RBW propagating in a constant
density environment for a parabolic GRB outflow. The dashed curve
represents the evolution with radius in the adiabatic case without
radiative drag. The red vertical lines represent the region where
the kinematic condition is satisfied, and the green line gives the
radius at which the dynamic condition starts to be satisfied. The
GRB parameters are given in the text.\label{fig2}}
\end{figure}

In a parabolic outflow the accumulated number of ambient particles
increases like $N(R)\propto R^{2}$, unlike the conic case where
$N(R)\propto R^{3}$. Therefore the evolution equation with the
previous assumption that $\dot{E} = \dot{E}_{\rm inj}$ are given by
\begin{equation}
\frac{dM}{dx} = 4\pi R_{0}^3 x \Gamma\rho - 4\pi x
R_{0}^3\rho(\Gamma - 1), \label{evop1}
\end{equation}
and
\begin{equation}
\frac{d\Gamma}{dx} = \frac{2\Gamma A}{3x(x^{2/3} + A)} - \frac{4\pi
x R_{0}^3 \rho \Gamma^2}{M} -
\frac{256\pi^2}{9M}\tau_{b}n_{e}\sigma_{T}x^2
R_{0}^4\rho\Gamma^4(\Gamma^2 - \Gamma). \label{evop2}
\end{equation}
Solving numerically these differential equations in a constant
density medium with the same GRB parameters used previously in the
conic case, gives the evolution of the Lorentz factor $\Gamma$ with
the radius $R$ as shown Figure \ref{fig2}. Again the dotted curve
represents the evolution in the adiabatic case without radiative
drag. In the parabolic configuration the expanding RBW accumulates
mass at a slower rate and therefore has a larger deceleration radius
$R_{D}$ than the conic case and for the same reason the luminosity
released is smaller.  In fact both threshold conditions are
satisfied during the acceleration phase of the RBW when the first
term on the RHS of Eq. (\ref{evop2}) is dominant, i.e., before the
RBW Lorentz factor reaches its asymptotic value $\Gamma = 200$.
These factors explain the delay and smoothness of the drop in
$\Gamma$ when the RBW becomes supercritical.

As can be seen in Figure \ref{fig2}, in this case too, the slowing
down of the RBW due to the radiative drag during the supercritical
phase, reduces $b\Gamma^5$ below its threshold value, thereby ending
the prompt GRB phase. This is followed by a period of constant
$\Gamma$, which is longer than that obtained previously, due to the
fact that the RBW becomes supercritical during its accelerated
expansion phase and the fact that for a given $E_0$ it sweeps-up
matter at a slower rate than a conical RBW. After all the
non-adiabatic effects have died down and once the RBW has
accumulated enough mass, the evolution of $\Gamma$ beyond the
deceleration radius follows the conventional decay $\Gamma(R)
\propto R^{-1}$ appropriate for a parabolic flow.

\subsection{The X-Ray Flux}

The fact that the Lorentz factor remains constant for a period does
not guarantee that the corresponding X-ray flux does too.
Such an outcome depends also on the particular process responsible
for the X-ray emission. For example, if the RBW propagates in a
medium of constant density (as assumed herein), with the magnetic
field in equipartition with the plasma (i.e. $B \simeq 0.4 (n_0/1
{\rm cm}^{-3})^{1/2} \Gamma$ G), the observed flux would increase
with time, because the source specific intensity would remain the
same (it sweeps the same amount of electrons per unit time) while
its solid angle (its size) increases. Therefore the constant XRT
emission during the ``plateau" stage imposes certain restrictions on
the emission process. If the observed X-ray emission is due to bulk
Comptonization, just like the $\gamma-$ray prompt emission, then the
$R^{-2}$ decrease of the ambient photons could indeed offset the
$\propto R^2$ increase of the RBW surface to produce a constant
X-ray flux. An alternative is that the ``plateau" X-ray emission is
due to synchrotron but by a magnetic field that decreases with
radius like $B(R) \propto R^{-1}$, since the emissivity is
proportional to $B(R)^2 \propto R^{-2}$, thereby offsetting again
the increase in the RBW area. Finally, if the ambient particle
density decreases like $R^{-2}$, i.e. the RBW propagates in a
stellar wind environment, that would also lead to a constant flux
\citep{Matzner12}; however, the present calculations for
the evolution of $\Gamma$ would have to be revised to reflect the
different density dependence on $R$.


%
%
The situation is not too different in the parabolic expansion case,
provided that the observer ``sees" only a fraction of the expanding
RBW front. In Figure 4 we show the bolometric light curve of a GRB
with the kinematic characteristics of the parabolic outflow shown in
Figure 2, assuming that the emission past the radiative reaction
slow down is due to bulk Comptonization, i.e. that the number of
photons decreases with radius like $R^{-2}$. One has to bear in mind
however, that this is only a toy model light curve, based on a
number of simplifying assumptions. However, this along with the
dynamics of radiation reaction feedback set the stage for a more
detailed future exploration of these issues.

\section{Summary and Discussion}

In the present work we have studied  the evolution of the Lorentz
factor $\Gamma$ of a GRB RBW from its origin (of $\Gamma_{0} \sim 1$
at $R = R_0$) through its acceleration ($\Gamma \propto R$ for conic
outflows and $\Gamma \propto R^{2/3}$ for parabolic outflows)
to saturation ($\Gamma = $ constant), and decay phases,
within the context of the GRB Supercritical Pile Model (SPM). We
have argued within the framework of well understood defined physical
processes and without the introduction of {\em a posteriori}
assumptions, that this evolution provides, among others: (a) A
definition of what constitutes the prompt GRB phase, its broader
spectral features and a criterion for its termination and of the
onset of the afterglow. (b) An account of the
steep-decline-and-plateau or the steep-decline-and-power law phase
of the GRB afterglow, observed in the largest number of afterglows
\citep{evans09} and their relation to the prompt emission
properties.

Of these, (a) has been discussed in several of our previous
publications \citep{kaz02,mas06,mas09} where it was shown that the
SPM can provide both an efficient, rapid conversion of the energy
stored in relativistic protons in the RBW of a GRB into electrons,
while at the same time producing a spectrum with $E_{\rm pk} \sim
m_ec^2$, in agreement with observations. In the present work this
process has been placed within the broader context of an evolving
RBW and it is shown that the prompt GRB phase lasts as long as this
efficient transfer of energy from protons into electrons is allowed
by the kinematics of the pair production process which depends
crucially on $\Gamma$. With the reduction of $\Gamma$ below the
critical value set by the $p\gamma \rightarrow p \, e^+ e^-$
reaction threshold, this transfer stops and the GRB luminosity
decreases precipitously along with the value of $E_{\rm pk}$.

Item (b) is the novel aspect of the present work, which shows that
incorporating the radiation reaction associated with the flowing of
the RBW through its upstream scattered radiation (or for that matter
any sufficiently intense ambient radiation) in the evolution of
$\Gamma$, can produce a small, sharp (over distance $\Delta R \lsim
R$) but important decrease of the RBW Lorentz factor value $\Gamma$.
Although small, this decrease is important because it pushes its
value below that of the kinematic threshold of the SPM (at least
within the confines of the simplified treatment of radiation
emission used herein). As a result the transfer of energy from
protons into electrons ceases and leads to a steep reduction in the
GRB luminosity by roughly a factor $\sim m_p/m_e$, consistent with
the decrease in luminosity between the prompt and afterglow GRB
luminosities. In figures 5a and 5b we present a sample of two such
XRT light curves, namely those of GRB 110420A and GRB 120213A, taken
from Nat Butler's compilation
(http://butler.lab.asu.edu/swift/older.html). In these figures we
note with the thick yellow arrow a range of $\simeq 2,000$ to
provide an eyeball estimate of the change in flux between the prompt
and afterglow stages, which indeed is consistent with the estimate
given here.  We have also estimated by inspection the same ratio in
a number of other bursts in the same list; we found some of them to
be smaller and some larger. Smaller values ensue in cases that not
all protons are ``burnt", as it happens to be the case in numerical
simulations of this process. Larger values of the ratio will be the
result of a significant decrease in $\Gamma$, considering that the
observed luminosity is proportional to $\sim \Gamma^4$. Such is the
case of figures 3a and 3b which produce respectively values
$\Gamma^4 \simeq 10 ~{\rm and}~ 100$. A more detailed
statistical analysis of this issue is currently under
consideration.

As long as the RBW radius at the point of this transition is smaller
than its deceleration radius $R_D$, following this  decrease in
$\Gamma$ (and $L$) due to the radiation reaction process, the
internal energy to rest mass ratio of the RBW is sufficiently high
to ensure that its Lorentz factor $\Gamma$ remains constant until
its radius reaches the value $R_D$, at which point it begins its
more conventional decline. We have also argued that during this
period of constant $\Gamma$, the value of the X-ray flux can remain
roughly constant, in agreement with observations, provided
that certain conditions are fulfilled as already described in
Section 3.3. Interestingly, synchrotron emission by a RBW
propagating into a medium of constant density and magnetic field
would yield a flux increasing with time and it is excluded in most
cases. The effect of radiation reaction feedback was considered in
\citet{mas08} and in more detail \citet{mas09}. However, the
conditions of the most detailed, latter study were such that the
radiation reaction transition radius was close to $R_D$, just like
that of Fig. 3b, so the constant $\Gamma$ section was not
discernible (nonetheless, the steep and then less steep decline of
the 10 keV flux was apparent in their figure 4).

Viewed more broadly, figures 1, 3a, 3b present a set of calculations
of the evolution of the Lorentz factor of three different RBWs with
the same initial conditions of radius, $R_0 = 10^{11}$ cm, internal
energy $E_0 = 10^{54}\ {\rm erg}$ and background (uniform) density,
$n = 100 \; {\rm cm}^{-3}$ but different values of their initial
internal rest mass-energy $M_0 c^2$, namely $M_{0}c^2 =
5\times10^{51},\ 3.5\times 10^{51},\ 2 \times 10^{51} \  {\rm erg}$
respectively. To these correspond the following values of asymptotic
Lorentz factor, $\Gamma_{\infty} = 200, 286, 500$. These values are
indeed achieved, and when both thresholds of the SPM are satisfied,
the proton energy is released to produce the main GRB emission and,
under the force of the radiation reaction, the Lorentz
factor is reduced sharply to $\Gamma \simeq 160$ in all three cases.
However, the duration of the plateau in the evolution of $\Gamma$
becomes shorter with a decreasing value of $M_0 c^2$. This is an
important fact in view of the systematics between the plateau X-ray
luminosity and its duration $T_{\rm brk}$ \citep{dai08,dai10}. The
goal of the present paper is not to resolve this issue but to point
out that the considerations discussed herein, which associate $R_D$
with $T_{\rm bk}$ rather than to the onset of the afterglow, provide
a novel framework within which such issues can be discussed and
perhaps resolved. Furthermore, since both emissions prior to and
post the radiation reaction reduction in $\Gamma$ take place while
$\Gamma$ was approximately constant, it may not be surprising to
find that GRB properties in these two stages are correlated, even
though one belongs to the prompt and the other to the afterglow GRB
stage. This is indeed the correlation found by \citet{SK12}.

One of the most challenging issues of the SPM, raised a number of
times at conferences and by the referee of the present note, is that
of the fast variability, $\Delta t \ll t$, of the GRB prompt
emission, a prominent characteristic of the majority of GRBs; this
poses a problem, considering that the propagation of blast wave in a
uniform medium cannot produce variability times shorter than
$R/c\Gamma^2$. Consideration of small, spherical inhomogeneities in
the swept-up medium was shown to be very inefficient \citep{sari97}.
The broadly accepted proposal as of this writing is that of
``internal shocks" \citep{rees94}, i.e. variations in the activity
of the GRB central engine that result in different sections of the
outflow catching-up with each other and colliding to produce the
observed variability. However, \citet{NK09} have argued that
this process has only a modest efficiency ($\sim 1-10 \%$); at the
same time they noted that model fits to the data of specific GRBs
obtain prompt emission locations (as does the SPM) much larger than
those inferred from the colliding shock locations. For these
reasons these authors opted for relativistic turbulence to provide
additional boosting to small regions within the broader blast wave
to indicate that this notion could resolve the GRB fast
variability issue. On the other hand, numerical simulations
by \citet{ZM13} have shown that such turbulence would quickly
dissipate in the absence of a continuous energy input. It is
possible that such an input is provided by the reconnection of the
turbulently amplified magnetic field in the post-shock region, as
suggested by \citet{ZY11}, who have presented an entire edifice of
GRB variability based on the notion of magnetic field reconnection.

While beyond the scope of the present work, one could
speculate that these processes may be possible to integrate within
the SPM.  Clearly, a uniform medium cannot produce the observed
variability and a collection of ``blobs" in the ambient medium
can only be efficient under the conditions discussed in
\citet{NK09}.  Nonetheless, recent particle-in-cell (PIC)
simulations \citep{silva03,nishikawa06,ruiz07} have shown that
relativistic shocks such as those of the GRB RBWs are unstable to
filamentation via the Weibel \citep{weibel59} or two-stream
\citep{buneman58} instabilities. If the column density of the RBW is
$\Sigma$, the splitting of the flow by these instabilities into,
say, $N$ filaments, will increase their local column density to
$\sim N \Sigma$ (there will likely be a distribution of columns,
each becoming supercritical at different times); since the crucial
quantity for the SPM is the local column, filamentation makes the
conversion of RBW kinetic energy to radiation all the more
efficient, with each such filament playing the role of the turbulent
eddies of the model of \citet{NK09}. Of course, each such filament
produces independently (via the SPM process) relativistic electrons
that plow through the ambient radiation to produce the observed
emission of what is perceived as a single pulse in the overall GRB
prompt emission. Admittedly, these PIC simulations probe
only scales associated with their microscopic plasma physics
quantities; however, \citet{silva03} do contend that these filaments
may eventually organize to much larger scale (of the order of the
shock width in our case). Whether these can reproduce the observed
GRB variability is, at present, an open issue. The duration of the
shots resulting from each such filament should then depend on the
local electron cooling time.

In summary, we would like note that the SPM, when
integrated within the entire evolution of a GRB blast wave,
provides, to the best of our knowledge, the first consolidation of
the broader temporal and spectral properties of both the
prompt and afterglow GRB stages within the framework of a single
model. Considering that this model involves essentially no free
{\em a priori} assumptions, it should not be expected to account for
specific features of specific bursts, rather it should be
considered as a broader framework within which one attempt to
account the more specific GRB systematics.  In this respect, we
find extremely encouraging the fact that the same physics
which provides for the efficient conversion of blast wave kinetic
energy to radiation and the value(s) of $E_{\rm pk}$, provides also an
account of the vexing XRT light curves. Furthermore, it makes a
prediction for the bolometric luminosity prior to and post the
radiation reaction reduction in $\Gamma$, indicating that this
should be of order $m_p/m_e$. Indeed a cursory search through the
combined BAT-XRT light curves, suggests that this is indeed the
case; we are currently involved in providing a more complete
statistic of the above statement which will appear in a future
publication.

\acknowledgments We would like to thank the anonymous referee for
their incisive, constructive comments which have added to the
completeness of this work. J.S. gratefully acknowledges financial
support from the University of Malta during his visit at NASA-GSFC
and the hospitality of the Astrophysics Science Division of GSFC.
D.K. acknowledges support by {\em Swift} and {\em Fermi} GO grants.

\clearpage

\clearpage

\clearpage

\begin{figure}
\begin{center}$
\begin{array}{cc}
\includegraphics[trim=0in 0in 0in
0in,keepaspectratio=false,width=3.0in,angle=-0,clip=false]{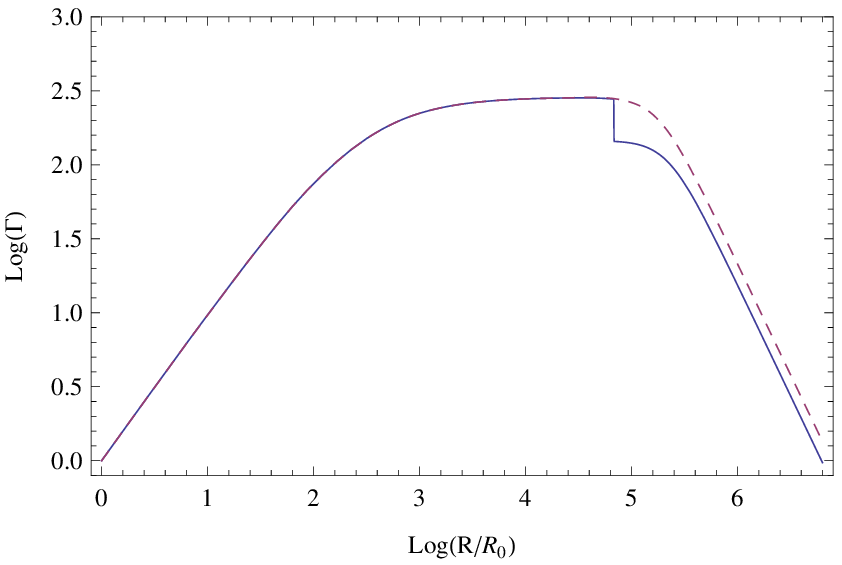}
\hskip 10pt
\includegraphics[trim=0in 0in 0in
0in,keepaspectratio=false,width=3.0in,angle=-0,clip=false]{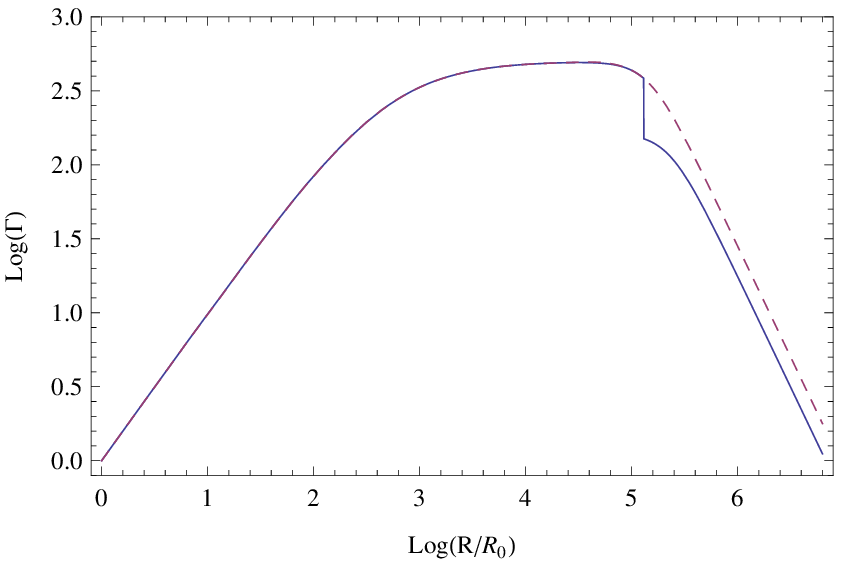}
\end{array}$
\end{center}
\caption{The Lorentz factor evolution of two conical RBWs
propagating in the constant density environment of figure 1. These
are distinguished from that of figure 1 only in their initial rest
mass energies which are $M_0c^2 = 3.5 \times 10^{51}, ~ 2.0 \times
10^{51}$ ergs for the left and right figures respectively. The
dashed curves represent the evolution with radius in the adiabatic
case without radiative drag. The rest of the GRB parameters are
given in the text.\label{fig3}}
\end{figure}

\clearpage

\begin{figure}
\includegraphics[scale=1.5]{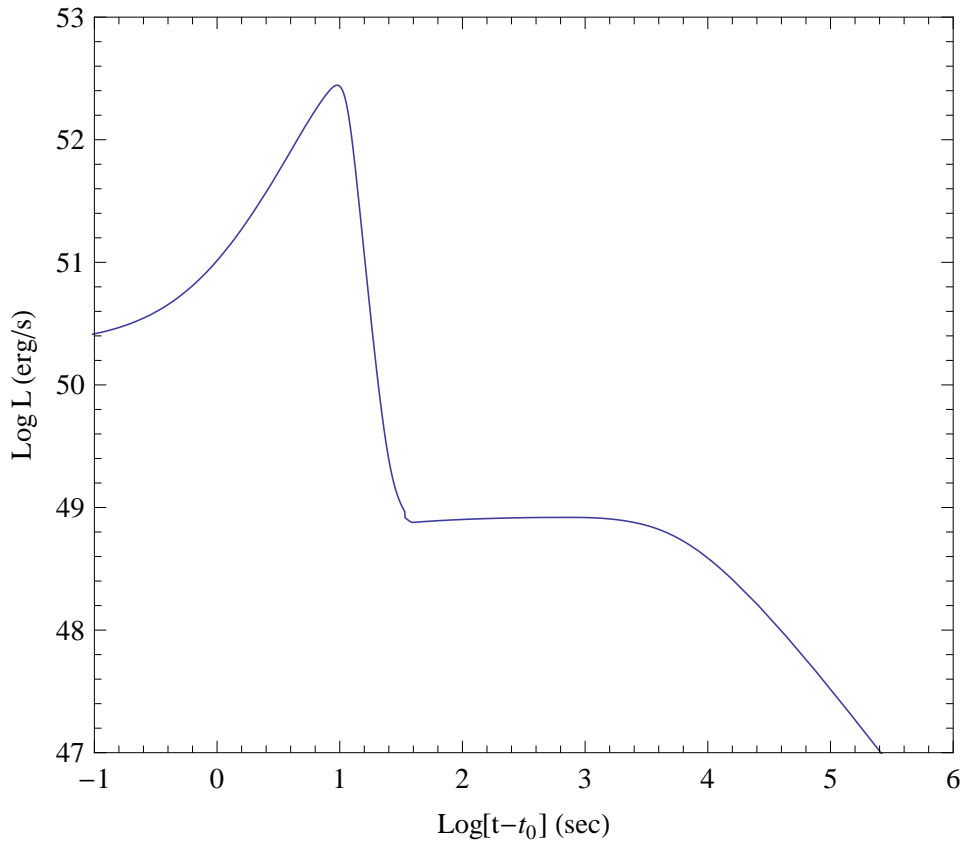}
\caption{The bolometric Luminosity of a RBW propagating in a
constant density environment for a parabolic GRB outflow.  The
rising part is the prompt BAT emission, while the steeply and
plateau one is the XRT afterglow one assuming the density of the
photons scattered in the ``mirror" decreases like $R^{-2}$. The time
$t_{0}$ represents the time of the onset of the burst.\label{fig4}}
\end{figure}

\clearpage

\begin{figure}
\begin{center}$
\begin{array}{cc}
\includegraphics[trim=0in 0in 0in
0in,keepaspectratio=false,width=3.0in,angle=-0,clip=false]{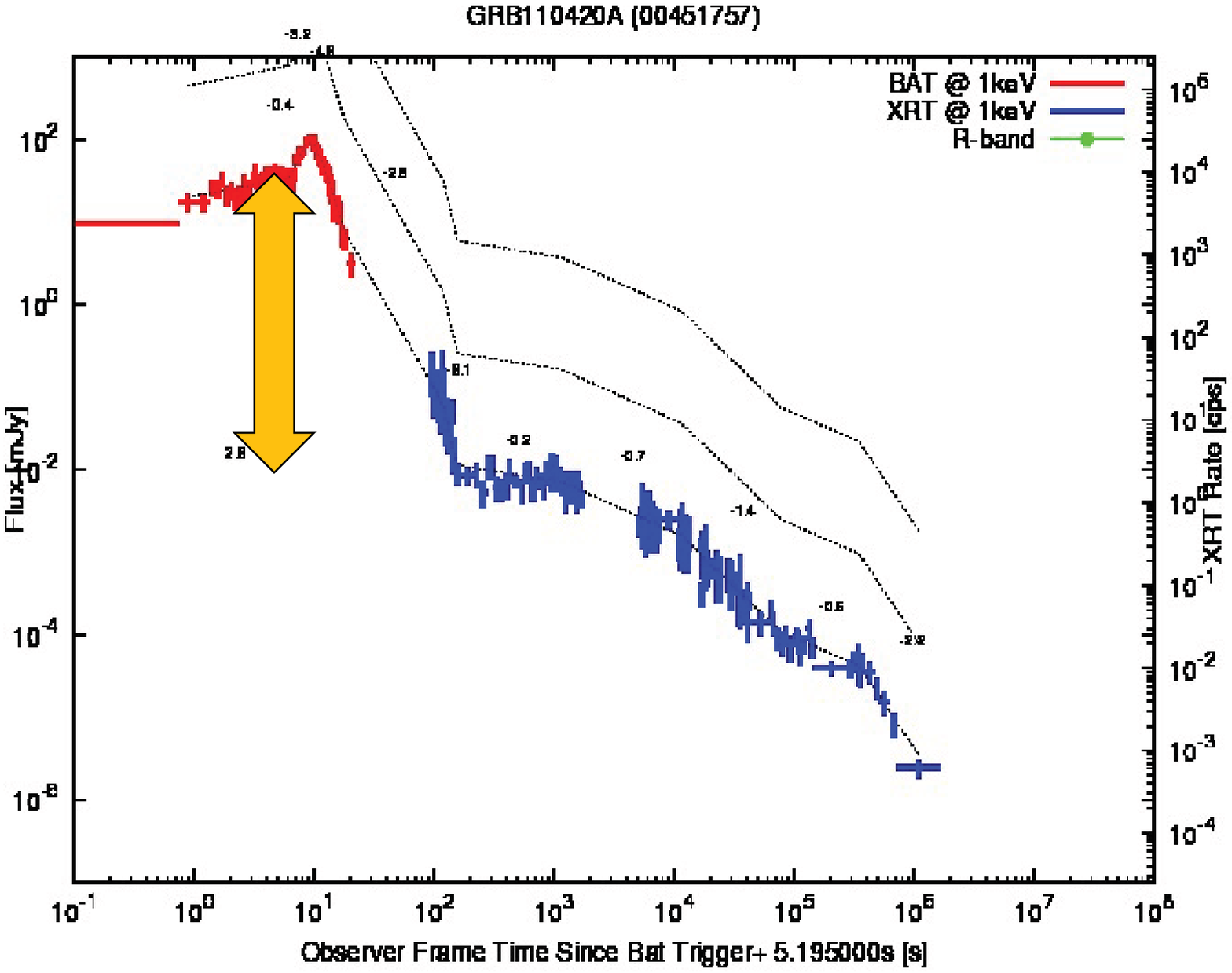}
\hskip 10pt
\includegraphics[trim=0in 0in 0in
0in,keepaspectratio=false,width=3.0in,angle=-0,clip=false]{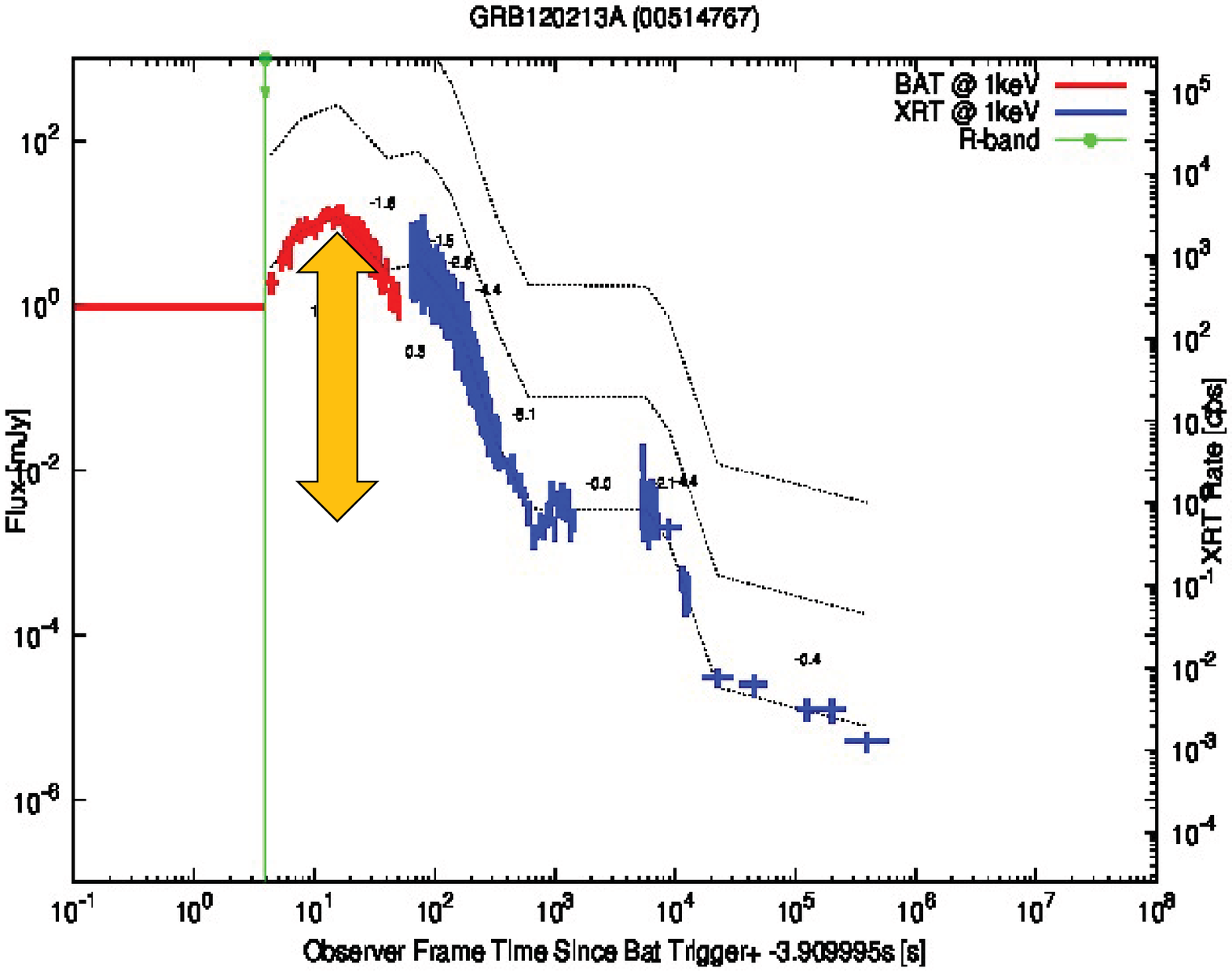}
\end{array}$
\end{center}
\caption{The flux evolution of the prompt ({\em Swift} BAT, red points)
and the afterglow ({\em Swift} XRT, blue points) of two GRBs, namely GRB110420A (left)
and GRB120213A (right). Both GRB exhibit the typical steep-decline-and-plateau shape
of afterglows discovered by XRT aboard {\em Swift}. The yellow arrow spans a
flux range of approximately $\simeq m_p/m_e$, the value implied by the physics
of the  SPM model.
 \label{fig5}}
\end{figure}

\end{document}